# Automatic Identification of Epileptic Seizures from EEG Signals using Sparse Representation-based Classification

S. Sheykhivand[1], T. Yousefi Rezaii[1], Z. Mousavi[2], A. Delpak[3] and A. Farzamnia[4]

[1]Biomedical Engineering Department, Faculty of Electrical and Computer Engineering, University of Tabriz, Tabriz, Iran
[2]Department of Mechanical Engineering, Faculty of Mechanical Engineering, University of Tabriz, Tabriz, Iran
[3]Medical Doctor, Research Assistant, Neurosciences Research Center, Tabriz University of Medical Sciences, Tabriz, Iran
[4]Assistant Professor, Faculty of Engineering, Universiti Malaysia Sabah, Kota Kinabalu, Sabah, Malaysia

Corresponding author: A. Farzamnia (e-mail:a@ieee.org) and T. Yousefi Rezaii (e-mail: yousefi@tabrizu.ac.ir)

**ABSTRACT** Identifying seizure activities in non-stationary electroencephalography (EEG) is a challenging task, since it is time-consuming, burdensome, and dependent on expensive human resources and subject to error and bias. A computerized seizure identification scheme can eradicate the above problems, assist clinicians and benefit epilepsy research. So far, several attempts were made to develop automatic systems to help neurophysiologists accurately identify epileptic seizures. In this research, a fully automated system is presented to automatically detect the various states of the epileptic seizure. The proposed method is based on sparse representation-based classification (SRC) theory and the proposed dictionary learning using electroencephalogram (EEG) signals. Furthermore, the proposed method does not require additional preprocessing and extraction of features which is common in the existing methods. The proposed method reached the sensitivity, specificity and accuracy of 100% in 8 out of 9 scenarios. It is also robust to the measurement noise of level as much as 0 dB. Compared to state-of-the-art algorithms and other common methods, the proposed method outperformed them in terms of sensitivity, specificity and accuracy. Moreover, it includes the most comprehensive scenarios for epileptic seizure detection, including different combinations of 2 to 5 class scenarios. The proposed automatic identification of epileptic seizures method can reduce the burden on medical professionals in analyzing large data through visual inspection as well as in deprived societies suffering from a shortage of functional magnetic resonance imaging (fMRI) equipment and specialized physician.

**INDEX TERMS** EEG, epilepsy, seizure, sparse representation-based classification, dictionary learning.

## I. INTRODUCTION

As reported by world health organization, about 50 million worldwide are suffering from epilepsy [1]. Epilepsy, as the second most common brain disorder after stroke, is characterized by an unexpected seizure, where, nerve cells generate abnormal electrical activities which leads to loss of consciousness in a limited period of time [2]. Proper diagnosis of epileptic seizure is essential to control and reduce the risk of epileptic attacks [3]. Currently, the diagnosis of epilepsy is based on neurological examination and auxiliary tests such as neuroimaging and Electroencephalography. EEG signals can reflect epileptic abnormalities between inter-ictal (between seizures) and ictal (during seizures) stages. Typically, neurons are in contact with each other by means of electrical potentials which follow a normal pattern in healthy human brain activity. While an abnormal electrical activity occurs in the brain's neural network during epilepsy, this incremental electrical activity can spread out through the entire cortex. A neurologist traditionally inspects the epileptic malformations. The interpretation of EEG signals using an intuitive evaluation is a time-consuming and tedious task, and the obtained results may vary and are limited according to the level of knowledge and expertise of the related physician. The use of anti-epileptic drugs have some restrictions and in 20-30% of patients is unable to control the seizure [3]. However, it is reported that using anti-epileptic drugs within pre-ictal stage might be more effective which prevents the occurrence of ictal stage and the possible physical damages caused by individual unconsciousness [3, 4]. Therefore, designing an automated computer diagnostic system seems to be essential to detect epileptic states from EEG signal based on machine learning techniques. In addition to helping the expert diagnose the epileptic stages, it will have the ability to continuously monitor the high-risk patients which alerts the seizure before its occurrence and inform the patient to take the drug. There are several stages of an epileptic seizure (brain activity of an individual with epilepsy), which play a major role in anticipating these seizures. Previous studies show that the seizure process is divided into four stages including pre-ictal, inter-ictal (pre-seizure disturbances), ictal (during a seizure), and postictal. Evidence suggests that seizures come from a recognizable brain state called pre-ictal, which can be considered as a clue to predict the upcoming stages (ictal) [4-6].

In the following the recent studies on the automatic identification of epileptic seizures are reviewed. Tzallas et al. [7] calculated the power spectrum density of EEG signal segment using a variety of time-frequency distributions and used PSD as a discriminative feature to classify epileptic seizure stages. Adeli et al. [8] reported a classification algorithm using wavelet transformation and nonlinear dynamics-based features such as the largest Lyapunov exponent and correlation dimension. Oweis et al. [9] extracted frequency features from the Hilbert-Huang transform. They also used the t-test to verify the importance of the features. The accuracy and specificity of their algorithm for classification of 2 epileptic and normal states were reported 94% and 96%, respectively. Bajaj et al. [10] used the empirical mode decomposition (EMD) to compute modulation bandwidth features and then utilized least squares-support vector machine (LS-SVM) for classifications. They also used the statistical test of Kruskal-Wallis to verify the features. The sensitivity, accuracy and specificity of their algorithm to classify 2 epileptic and normal states were reported 100%, 99% and 99% respectively. Alam et al. [11] used EMD and artificial neural networks (ANN) for the identification of epilepsy. Both the above methods are affected by mode-mixing problems due to the use of EMD, meaning that EMD may result in varying oscillations in the same mode or similar oscillations in different modes. Peker et al. [12] extracted five statistical features using dual-tree complex wavelet transform and then applied complex-valued neural network transformations to classify epileptic seizure states in 4 different scenarios. They also used a 10-fold cross validation to evaluate their algorithm. Wang et al. [13] introduced an autoregressive multivariate, partially directed coherence and SVM classification for the automatic seizure detection. Samiee et al. [14] proposed a rationally discreet short-time Fourier transform and statistical features for the classification of epileptic seizures. Das et al. [15] employed normal inverse Gaussian parameters in the wavelet domain into their seizure classification scheme. Guler et al. [16] proposed a seizure detection scheme using wavelet coefficients and a multi-class support vector machine based on the Lyapunov exponents. Guo et al. [17] presented a seizure detection model using the line length features of EEG wavelet sub-bands, followed by an artificial neural network for classification. Swami et al. [18] have extracted features such as energy, Shannon entropy and few other statistical features from EEG sub-bands and feed them to a general neural regression network classifier. Hassan et al. [19] presented an automatic diagnostic design for various epileptic seizures based on the tunable-Q wavelet transformation and bootstrap classification leading to an accuracy of 99%. Sharma et al. [20] used flexible analytical time-frequency wavelet transformation and calculated fractal dimensions to discriminate various epileptic states. They have reported accuracy of 99% for their proposed method based on LS-SVM classifier. Acharya et al. [21] proposed conventional neural networks (CNN) for automatic identification of pre-ictal, inter-ictal and normal states from EEG signal. The propsed CNN architecture includes 10 convolution and 3 fully connected layers, which lead to accuracy and sensitivity of 88% and 95%, respectively.

The challenging step in automatic epileptic seizure detection is to select the discriminative features of various stages of epilepsy. In the majority of the existing works, at first different time, frequency, time-frequency as well as statistical features are extracted, and then the best discriminatory features are picked either manually or using traditional feature selection methods, which is a time-consuming procedure demanding high computational complexity. In addition, the best features in one case/subject may not be considered optimal in another. Therefore, it is crucial to implement an algorithm which learns the appropriate features corresponding to each case/subject. This will remain as the main advantage of this paper. At first, a sparsifying transform is introduced for the EEG signal of each designated state of epileptic seizure. Then, the proposed online dictionary learning is used to obtain the sparsest representation for each of the states and sparse representation-based classification (SRC) is applied in order to identify different classes. The proposed approach can be considered as an end-to-end classifier, in which there is no need to a feature selection/extraction procedure and the discriminative features of each class will be automatically learned during dictionary learning. In dictionary learning, there are two parameters which need to be optimized, namely, the atoms of the dictionary and the sparse coefficients that relate the atoms of the dictionary to the training data set. Since the dictionary learning problem is NP-hard, dictionary learning algorithms use alternating methods to optimize the parameters. In the first step, called sparse coding, the sparse coefficients are calculated by considering a pre-defined dictionary. The most conventional algorithms used as the first step are Matching Pursuit (MP), OMP [22, 23]. In the second step, the sparse coefficients that are calculated in the previous step are used to update the atoms of the dictionary. These two steps are repeated until the dictionary learning algorithm converges.

The most of the attention in dictionary learning problem is to improve the algorithms used in the second step. Some of the important algorithms that are used in this step are: Method of Optimal Directions (MOD) [24], Recursive Least Squares (RLS) dictionary learning [25], Online Dictionary Learning (ODL) for sparse representation [26] and K-Singular Value Decomposition (K-SVD) method [27].

In this paper, we have also focused on various scenarios for occurrence of epileptic seizure considered in the related literature (and also the existing datasets) and evaluated the proposed algorithm in 9 most complex scenarios to identify the specific states related to the epileptic seizure. We also presented a proposed algorithm for learning the dictionary for each class. The results very promising, such that in 8 out of 9 scenarios the classification accuracy was 100% while in the remaining one, it was as much as 95%.

Finally, unbalanced class data is another challenging issue in the previous work, where, the authors used data augmentation methods to make the data from different classes balanced, or some classifiers which are not sensitive to unbalanced class data, while, the proposed method for dictionary learning in this paper is almost insensitive to unbalanced class population.

The remaining of the paper is organized as follows: The used database and the related mathematical background of SRC are given in Section 2. Theory of the proposed algorithm is discussed in Section 3. The simulation results and comparison of the proposed method with the state-of-the-art are given in Section 4, followed by the conclusion remarks in Section 5.

## II. MATERIALS AND METHODS

In this section, we first introduce the EEG database from the University of Bonn. Then, the mathematical background of SRC theory will be provided.

### A. EEG DATABASE

In this paper, we have used the EEG database created by Andrzezak et al. [6] at the University of Bonn. This database is widely used in seizure detection techniques which is publicly available. It consists of 500 single-channel EEG signal epochs in 5 subsets (A, B, C, D, and E) from both normal and individuals suffering from seizure (100 epochs from each subset). Sample EEG epochs belonging to the subsets; A, B, C, D, and E is shown in Fig. 1. Subsets A and B contain EEG data, recorded in a relaxed and awake state from five healthy subjects with open eyes (subset A) and closed eyes (subset B). The subsets C and D were recorded in five patients who had complete seizure control after epileptic focus resection. The EEG signals in subset C were recorded from the formation of the opposite brain hemisphere (inter-ictal), while the signals in D were recorded from the hippocampal formation identified as an epileptogenic area. Finally, subset E contains only ictal activity in the epileptogenic area. All subsets include 100 EEG segments, whereas each segment has a sampling rate of 173.610 Hz for 23.6 seconds (thus containing 4097 samples).

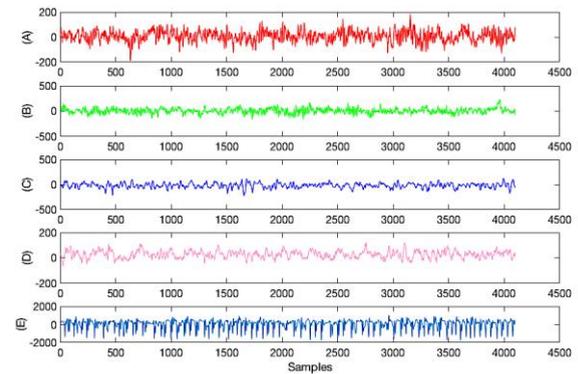

**FIGURE 1.** Sample EEG epochs belonging to the subsets; A, B, C, D, and E.

### B. SPARSE REPRESENTATION-BASED CLASSIFICATION

In the following, the mathematical background of SRC algorithm is introduced. The main idea in SRC is to obtain a sparsifying transform for each of the classes using training data set and then classify the data from test set based on the residual reconstruction error of the test data using each of the sparsifying transforms [28]. In mathematical terms, a signal $y \in \mathbb{R}^N$ is called $k$-sparse if at most $k$ out of $N$ samples are nonzero (this is also stated as $y_0 \leq k$, where $\|\ \|_0$ is the zero norm of vector $y$). Most of the existing natural signals including EEG are sparse or have sparse representation in a specific domain (transform). Considering $\phi \in \mathbb{R}^{N \times M}$ as the sparsifying dictionary, the sparse representation of the data signal vector $y$ can be obtained by solving the linear system of equations $y = \phi x$. Gathering length $N$ data vectors of class $i$ from $S$ EEG recording electrodes in the columns of a single matrix $Y^i$, the sparse representation model for multi-electrode EEG signal can be obtained as follows:

$$Y^i = \phi^i X^i, \quad i = 1,...,C \qquad (1)$$

Where $C$ is the total number of classes, $Y^i = \left(y_1^i, y_2^i, ..., y_S^i\right) \in \mathbb{R}^{N \times S}$, and $X^i = \left(x_1^i, x_2^i, ..., x_S^i\right) \in \mathbb{R}^{M \times S}$ is the corresponding sparse representation. Now, assuming the test data sample $Y$, the corresponding sparse representation will be obtained by solving the following optimization problem using the dictionaries of each class, to obtain $\alpha_i$:

$$z_j^i = \min_{\alpha} \alpha_1 \quad \text{s.t.} \quad y_j^i = \phi^i \alpha, \quad j = 1,...,S \text{ and } i = 1,...,C \qquad (2)$$

Where $z_j^i$ is the sparse representation of the j-th column of the test data matrix, i.e., $y_j$, using the sparsifying dictionary of class $i$, $\phi^i$. Finally, SRC classifies the data by comparing the residual error of the reconstructed EEG signal using the dictionaries of all classes, i.e.,

$$i^* = \underset{i=1,...,C}{\operatorname{argmin}}\, r_i(Y) = \left\| Y - \phi^i Z^i \right\|_F \quad (3)$$

where $Z^i = (z_1^i, z_2^i, ..., z_S^i)$, $\|\ \|_F$ is the Frobenius norm and $i^*$ is the estimated label of the test data. In many practical cases, however, the test data are accompanied by some bounded observation/measurement noise, where the optimization problem in (2) can be restated as follows in order to account for the noise component:

$$z_j^i = \min_{\alpha} \alpha_1 \text{ s.t. } \left\| y_j^i - \phi^i \alpha \right\|_2 \leq \varepsilon,\ j=1,...,S \text{ and } i=1,...,C \quad (4)$$

$\varepsilon$ is a positive and small number that corresponds to the noise energy.

## II. THE PROPOSED METHOD VIA DICTIONARY LEARNING AND SPARSE REPRESENTATION-BASED CLASSIfiCATION

In this section, the proposed method to automatically classification of epileptic seizure states is described. The block diagram of the proposed method is shown in Fig. 2. In the first phase, the recorded signals are divided into two subsets of test and training data (data collection). In the second phase, the dictionary matrices are updated for the different classes using the training data (dictionary learning). The sparse representation of the test data is obtained in the third stage using the dictionary matrices from the dictionary learning phase and then, they are reconstructed (reconstruction phase). Finally, in the fourth phase, automatic identification of epileptic seizures is performed based on the difference between initial (original) and the reconstructed signals from the third stage (classification phase). In the upcoming subsections, at first, online dictionary learning algorithm is discussed followed by the introduction of the proposed classification procedure and its parameters.

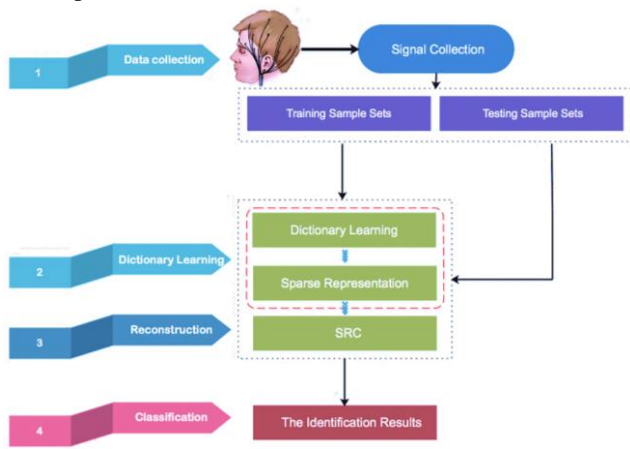

**FIGURE 2.** The block-diagram of the proposed method.

### A. THE PROPOSED DICTIONARY LEARNING

In general, the dictionary is referred to a set of atoms (columns of the dictionary matrix), which can be used to represent an underlying data as a linear combination of its atoms. Dictionaries which are used to obtain sparse representation for the signals are called sparsifying dictionaries and divided into two categories of deterministic and training-based dictionaries. Deterministic sparsifying dictionaries are not dependent on the underlying signal, like FFT and DCT bases matrices, while the entries of the training-based sparsifying dictionaries are completely dependent on the signal to be represented. Training-based dictionaries are signal-specific and can obtain the sparsest representation of a specific signal. Dictionary learning algorithms use training data in two manners: batch learning methods and sequential learning methods. In batch learning, the whole training data is used at once in order to obtain the atoms of the sparsifying dictionary. This method often has high computational burden, while the sequential methods in which the training data is utilized in a sequential manner have relatively lower computational burden. In online dictionary learning (a kind of sequential learning), starting from an initial solution/guess for the dictionary, its atoms are updated in a recursive manner as the new training data becomes available. In this paper, a new online dictionary learning algorithm, namely, correlation-based weighted recursive least square update (CBWRLSU), is proposed to update the atoms of the dictionary one by one based on their correlation with the new training data. This method has two major advantages: First, it significantly reduces the computational burden of heavy matrix-inversion by reducing the dimension of the matrix, which should be inverted. Second, it prevents the updating of the unnecessary atom. Algorithm 1 shows the summary of CBWRLSU dictionary learning.

In the proposed method, instead of the forgetting factor, a new data correlation with the previous data is used and the data correlated with the new data is used to update the atoms of the dictionary as well as the new data from the correction coefficient given in [29].

### B. SRC USING PROPOSED CBWRLSU DICTIONARY LEARNING

First of all, for the collected signals of epileptic seizure states, the over-complete learned dictionary from training samples for the state $i(i=1, 2, ..., C)$ using CBWRLSU algorithm is denoted as $\phi_i$. Then, the sparse representation for a test data $y$ (of unknown label) will be obtained using all of the $C$ learned dictionaries, leading to their corresponding sparse representations as $X_i$, $i=1, 2, ..., W$. The reconstruction error for the test data $y$ using the sparsifying dictionary from $i$-th state, i.e., $e_i$, can be calculated as:

$$e_i = \left\| y - \phi_i x_i \right\|_2^2 \quad (5)$$

| | |
|---|---|
| **Algorithm 1**: Proposed CBWRLSU dictionary learning algorithm method | |
| 1. | Initialize $\phi$ and $C$ |
| 2. | For ($i = 1: L$) |
| 3. | Get the new training data $y_i$ |
| 4. | Find $x_i$, sparse representation of $y_i$, using OMP |
| 5. | Find $\Omega(y_i)$, indices of previous signals which use common atoms in their sparse representation with $y_i$ |
| 6. | Find $Y'(y_i) \in \mathbb{R}^{m \times q_i}$, the set of all previous signals correlated with $y_i$ |
| 7. | Find $\phi'(y_i)$, the subset of $\phi$ which deals with $Y'(y_i)$ |
| 8. | For ($j = 1: q$) |
| 9. | Calculate $u_j(y_i) = C_{j-1}^{-1}(y_i) X_j(y_i)$ |
| 10. | Calculate $e_j(y_i) = Y'_j(y_i) - \phi'_{j-1}(y_i) X'_j(y_i)$ |
| 11. | Calculate $\omega_j(y_i)$, the correction weight using $$Y'_j(y_i) = \left[ Y'_{j-1}(y_i), \sqrt{\omega_j(y_i)} Y'_j(y_i) \right]$$ $$X'_j(y_i) = \left[ X'_{j-1}(y_i), \sqrt{\omega_j(y_i)} X'_j(y_i) \right]$$ |
| 12. | Calculate step size $\beta_j$ using $$u_{j+1}(y_i) = C_j^{-1}(y_i) X'_{j+1}(y_i), \ u_{j+1}^T(y_i) = X'^T_{j+1}(y_i) C_j^{-1}(y_i)$$ |
| 13. | Update $\phi'_j(y_i)$ using $$Y'_{j+1}(y_i) = \phi'_j(y_i) X'_{j+1}(y_i) = B_j(y_i) C_j^{-1}(y_i) X'_{j+1}(y_i)$$ $$e_{j+1}(y_i) = Y'_{j+1}(y_i) - \phi'_j(y_i) X'_{j+1}(y_i) = Y'_{j+1}(y_i) - \hat{Y}'_{j+1}(y_i)$$ and normalize its columns |
| 14. | Update $C_j^{-1}(y_i)$ for next step using $$\phi'_{j+1}(y_i) = \phi'_j(y_i) + \beta_{j+1} e_{j+1}(y_i) u_{j+1}^T(y_i)$$ |
| 15. | end |
| 16. | Replace the updated atoms of $\phi'_j(y_i)$ into the original dictionary $\phi$ |
| 17. | Update sparse coding of $y_i$ using OMP |
| 18. | end |

Finally, the data will be assigned a label, $j^*$, based on the solution of the following optimization problem:

$$j^* = \underset{i=1,\dots,C}{\operatorname{argmin}} e_i \qquad (6)$$

This procedure is depicted in Fig. 3.

The trial and error procedure is followed to determine the parameters of the proposed method. Since the length of each segment of is considered to be equal to the length of the sample data (4097 samples), the dimensions of the sparsifying dictionary is set to 4097×6000. In the training and testing processes, 90% of the data is randomly used for training and the remaining 10% for testing and 10-fold cross-validation is used to evaluate the classifier. The sparsity parameter k is empirically set to 10 for both learning and classification procedures.

## IV. SIMULATION RESULTS

The simulation results of the proposed method are presented in this section. The simulations are conducted on a PC with 8 GB of RAM and a 1.6 GHz core i5 CPU. In order to assess the classification performance of the proposed algorithm in different scenarios in terms of complexity as well as clinical relevance, nine different scenarios (namely case I to IX in Table. 1) were considered based on different combinations of the five existing EEG subsets (A, B, C, D and E) introduced in Section 2.1. These cases consist of four 2-class, three 3-class, and one 4 as well as one 5-class problems, constituting a more practical as well as a fair testbed to compare with the existing state-of-the-art. In order to visually asses the reconstruction performance of the proposed algorithm, a random sample is picked from each of the subsets and the original and reconstructed signals are plotted in Fig. 4, which shows that the reconstructed signals are quite consistent with the original ones. In order to gain more insight, the reconstructed signals of 90 samples of each subset (for training dataset) are shown in Fig. 5 at a particular time instance. Furthermore, as a quantitative measure for the reconstruction performance, the normalized reconstruction error ($E = y - \hat{y} / y$), for the segment of the signal in Fig. 5 is computed and plotted in Fig. 6. Accordingly, it can be concluded that the samples could be efficiently encoded as sparse representations using learned atoms. To put it more clearly, we chose one test sample from each subset (A, B, C, D and E) and the sparse representation coefficients of these five test samples based on their corresponding learned dictionaries are given in Fig. 7.

In terms of the computational complexity of the dictionary learning procedure, the runtime of the proposed algorithm for training each dictionary using the corresponding training dataset is roughly 28 minutes. In other words, a total of 140 minutes was spent on training 5 dictionaries (for each subset), while only 6 seconds were spent on classifying the total testing dataset given the trained dictionaries.

TABLE I
NINE DIFFERENT CLASSIFICATION CASES CONSIDERED IN THIS STUDY AND THEIR DESCRIPTION.

| Case | Class | Account |
|---|---|---|
| I | E-A | Seizure and Healthy (eyes-open) |
| II | E-B | Seizure and Healthy (eyes-closed) |
| III | E-B-D | Ictal, Healthy (eyes-closed) and Inter-Ictal |
| IV | E-C | Ictal and Inter-Ictal |
| V | E-D | Ictal and Inter-Ictal |
| VI | AB-E | Seizure and Healthy (eyes-open and eyes-closed) |
| VII | E-CD | Ictal and Inter-Ictal |
| VIII | AB-CD | Healthy (eyes-open and eyes-closed) and Inter-Ictal |
| IX | ABCD-E | Healthy (eyes-open and eyes-closed) with Inter-Ictal and Ictal |

In order to evaluate the classification of the proposed method for 9 different predefined cases, the classification performance in terms of accuracy, sensitivity and specificity is shown in Table. 2. It is evident from Table. 2 that among various clinically important cases, maximum accuracy, sensitivity and specificity for 8 out of 9 predefined cases is obtained, which is 100 percent, while the accuracy, sensitivity and specificity for the remaining VIII case is still very promising.

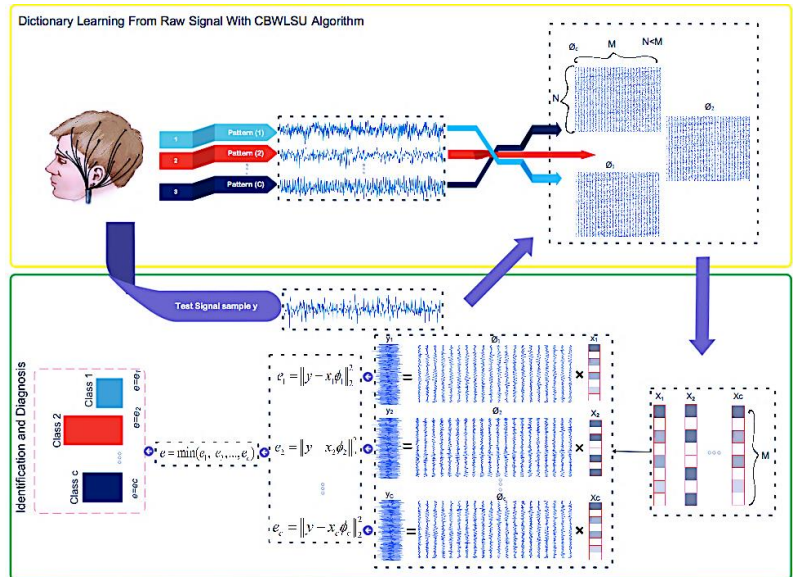

**FIGURE 3.** Block Diagram for automatic identification of epileptic seizures.

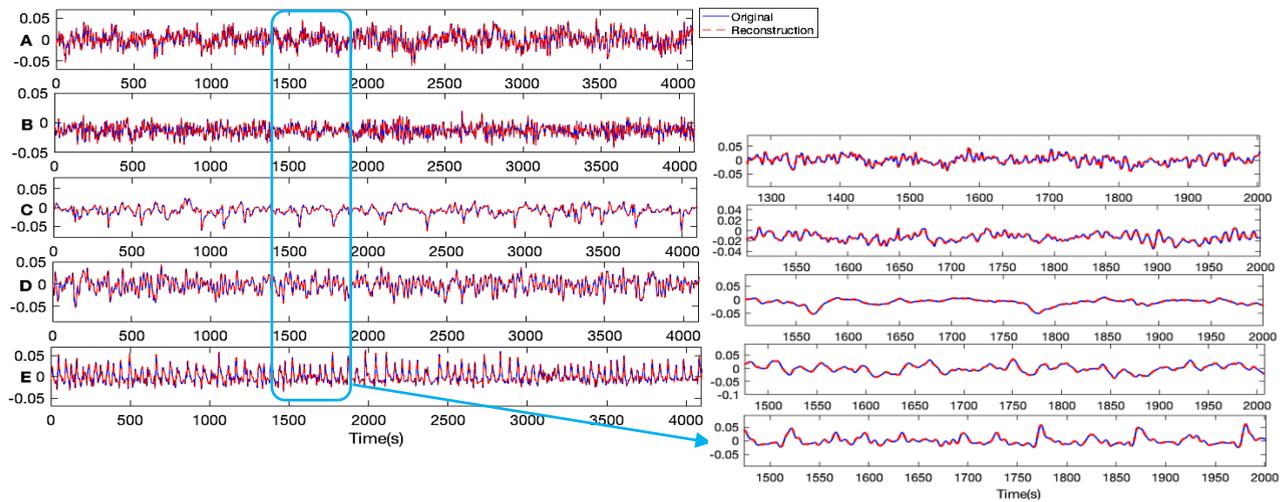

**FIGURE 4.** Original and reconstructed signals for each subset (A, B, C, D and E) for the sample no 50.

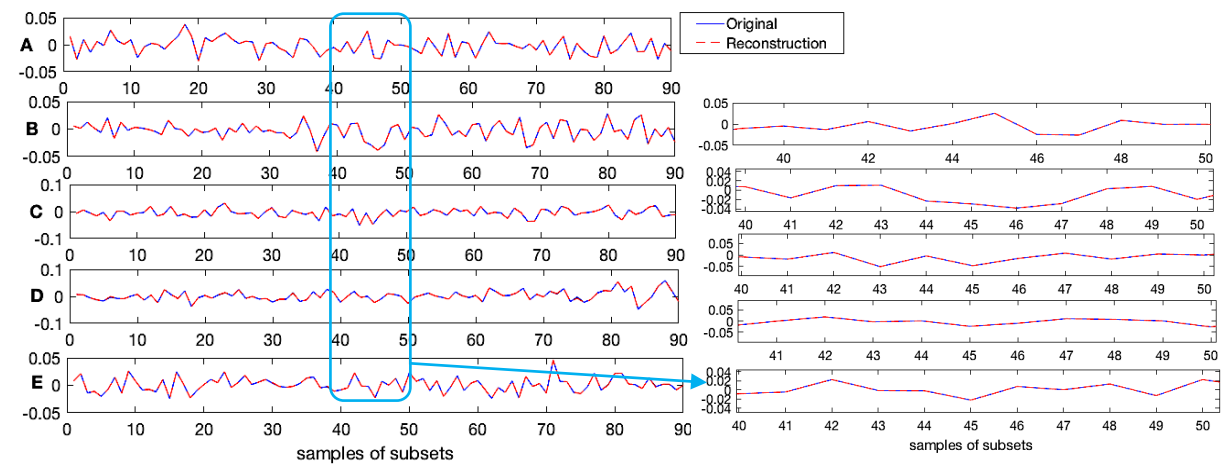

**FIGURE 5.** 90 samples of the reconstructed signals (for training dataset) at a particular time for each subset.

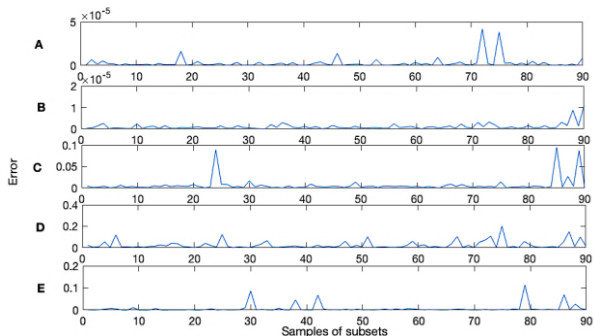

**FIGURE 6.** Reconstruction error ( $E = y - \hat{y} / y$ ) for the samples of the subsets in Fig. 5.

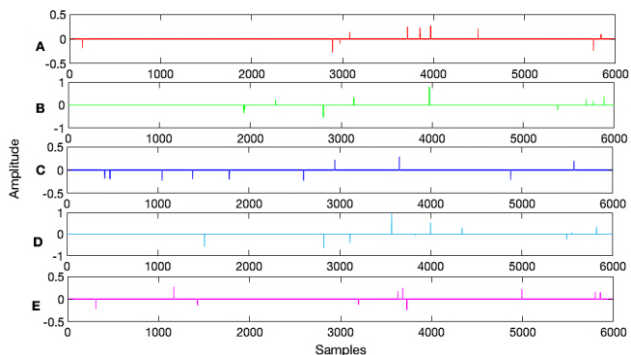

**FIGURE 7.** 90 samples of the reconstructed signals (for training dataset) at a particular time for each subset.

During recent years, several automatic seizure detection methods using EEG signal were proposed. In Table. 3, we compared various studies conducted on the same database to classify different predefined cases using EEG signals. The best results are highlighted in boldface. It is clear from Table. 3 that our proposed method offers the highest accuracy, sensitivity, and specificity for all 9 cases among all the comparative methods. In previous studies, common methods such as WT, EMD, etc. were used to extract the important characteristics and features of the signal, involving some common problems regarding the parameters of the feature selection/extraction procedure such as choosing the type of the mother wavelet, the number of decomposition levels, and etc. One of the most important advantages of the proposed method compared with the other methods is that the feature extraction is automatically done based on dictionary learning and no feature selection procedure is needed.

To illustrate the performance of the proposed CBWRLS method with various data types as input, the classification accuracy is obtained using the other common methods for 3 different predefined cases (I, III and VIII). In this regard, time data and several manual features from time data along with BPNN and SVM are selected as the comparative methods [39-42]. The Gaussian Radial Basis Function (RBF) is used as the kernel function of the SVM, and the grid search method is used to optimize the kernel parameters. In order to achieve better results from the BPNN model, the number of layers and hyper-parameters are adjusted by different data types. The parameters of the minimum, maximum, skewness, crest factor, variance, root mean square (RMS), mean and kurtosis are chosen as the manual features of the time domain (time features). [43, 44]. The testing accuracy of the different methods based on the feature learning from raw data and the manual features are presented in Table 4, where the result of the proposed CBWRLSU method is marked in bold.

TABLE II
NINE DIFFERENT CLASSIFICATION CASES CONSIDERED IN THIS STUDY AND THEIR DESCRIPTION.

| Case | Accuracy (%) | Sensitivity (%) | Specificity (%) |
|------|--------------|-----------------|-----------------|
| I    | 100          | 100             | 100             |
| II   | 100          | 100             | 100             |
| III  | 100          | 100             | 100             |
| IV   | 100          | 100             | 100             |
| V    | 100          | 100             | 100             |
| VI   | 100          | 100             | 100             |
| VII  | 100          | 100             | 100             |
| VIII | 95           | 95.45           | 95              |
| IX   | 100          | 100             | 100             |

Comparing the performance of feature learning and manual features, feature learning from raw time data with the proposed CBWRLS method provides better results than manual features. This result is significantly correlated with the unique Algorithm 1 of the proposed CBWRLS method, which can automatically extract the useful features for classification. While proposed CBWRLS has a better result with feature learning from raw time data, all the tested models, including CBWRLS, BPNN and SVM provide similar results with manual features. This indicates that the CBWRLS cannot achieve much more improvements in the identification of epileptic seizures than traditional methods without the ability of feature learning.

In order to assess the performance of the proposed method against observation noise, white Gaussian noise of SNR -20 to 20 dB is added as the measurement noise to the EEG signals and the classification accuracy for all 9 cases is reported in Fig. 8.

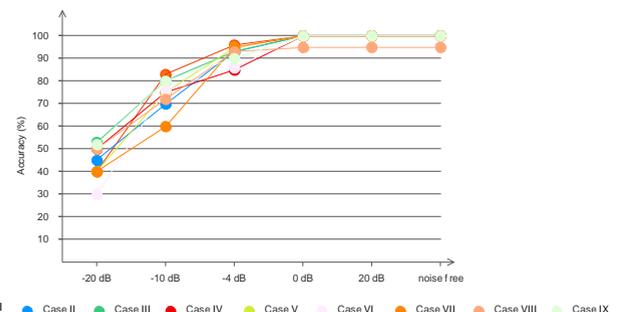

**FIGURE 8.** Accuracy of the proposed method versus SNR in additive white Gaussian noise scenario.

TABLE III
THE PERFORMANCE OF THE PROPOSED METHOD COMPARED WITH THE OTHER METHODS ON THE BONN EEG DATABASE.

| Case | Authors | Accuracy (%) | Sensitivity (%) | Specificity (%) |
|---|---|---|---|---|
| I | Guo et al. [17] | 95.20 | 98.17 | 92.12 |
| | Polat et al.* [30] | 98.72 | 99.40 | 99.31 |
| | Acharya et al. [31] | 99 | 99 | 99 |
| | Subasi [32] | 94.5 | 95 | 94 |
| | Tzallas et al.* [7] | **100** | **100** | **100** |
| | Orhan et al.* [33] | **100** | 100 | 100 |
| | Nicolaou et al. [34] | 99.50 | - | - |
| | Kaya et al.* [35] | 98 | 99 | 97 |
| | Peker et al.* [12] | **100** | **100** | **100** |
| | Samiee et al. [14] | 99.80 | 99.6 | 99.9 |
| | Swami et al.* [18] | **100** | - | - |
| | Hassan et al. [19] | **100** | **100** | **100** |
| | Sharma et al.* [20] | **100** | **100** | **100** |
| | **Proposed Method*** | **100** | **100** | **100** |
| II | Nicolaou et al. [34] | 82.88 | - | - |
| | Samiee et al. [14] | 99.30 | 99 | 99.6 |
| | Swami et al.[18] | 98.89 | - | - |
| | Sharma et al.* [20] | **100** | **100** | **100** |
| | **Proposed Method*** | **100** | **100** | **100** |
| III | Guo et al. [17] | 93.5 | - | - |
| | Bhattavh et al. [36] | 98.6 | - | - |
| | Acharya et al* [21] | 87.7 | 95 | 90 |
| | **Proposed Method*** | **100** | **100** | **100** |
| IV | Nicolaou et al.[ 34] | 88 | - | - |
| | Samiee et al. [14] | 98.50 | 99.3 | 97.7 |
| | Swami et al.* [18] | 98.72 | - | - |
| | Hasan et al. [19] | **100** | **100** | **100** |
| | Sharma et al.* [20] | 99 | **100** | 98 |
| | **Proposed Method*** | **100** | **100** | **100** |
| V | Nicolaou et al. [34] | 79.94 | - | - |
| | Kaya et al.* [35] | 95.50 | 96 | 95 |
| | Siuly et al. [37] | 93.60 | 89.40 | 97.80 |
| | Kumar et al. [38] | 93 | 94 | 92 |
| | Alam et al. [11] | **100** | - | - |
| | Hassan et al. [19] | **100** | **100** | **100** |
| | Sharma et al.* [20] | 98.50 | **100** | 97 |
| | **Proposed Method*** | **100** | **100** | **100** |
| VI | Swami et al. [18] | 99 | - | - |
| | Sharma et al. * [20] | **100** | **100** | **100** |
| | **Proposed Method*** | **100** | **100** | **100** |
| VII | Kaya et al.* [35] | 97 | 98 | 95 |
| | Swami et al.* [18] | 95.15 | - | - |
| | Hassan et al. [19] | 98.67 | 98.67 | 98.67 |
| | Sharma et al.* [20] | 98.67 | **100** | 96 |
| | **Proposed Method*** | **100** | **100** | **100** |
| VIII | Sharma et al.* [20] | 92.50 | 90.50 | 94.50 |
| | **Proposed Method*** | **95** | **95.45** | **95** |
| IX | Orhan et al.* [33] | 99.60 | 98.04 | **100** |
| | Guo et al. [17] | 97.77 | 98.61 | 94.60 |
| | Hassan et al. [19] | 99.60 | 99.49 | 100 |
| | Sharma et al.* [20] | 99.20 | **100** | 96 |
| | **Proposed Method*** | **100** | **100** | **100** |

(*Using 10-fold cross-validation)

TABLE IV
THE TESTING ACCURACY OF DIFFERENT METHODS FOR IDENTIFICATION OF EPILEPTIC SEIZURES FOR 3 DIFFERENT PREDEFINED CASES (I, III AND VIII).

| Methods | Feature learning from raw data | | | Manual features | | |
|---|---|---|---|---|---|---|
| | Case I | Case III | Case VIII | Case I | Case II | Case VIII |
| **SVM** | 89.7% | 92.2% | 81.9% | 95.4% | 94.1% | 91.9% |
| **BPNN** | 94.3% | 94% | 88.4% | 98.7% | 96% | 93.2% |
| **P-M (CBWRLSU)** | **100%** | **100%** | **95%** | 97% | 95.3% | 92.2% |

As it is seen, the classification performance of the proposed method is considerably robust to the measurement noise in a wide range of SNR, such that the accuracy is still more than 80% for SNR of -4 to 20 dB. Despite the contributions, this work has some limitations, as with other previous studies. First, notwithstanding the use of the Bonn database, a clinical validation study based on a bigger dataset is still necessary. Second, the training time of the proposed algorithm is relatively high, which can be solved using graphical processing unit (GPU) systems.

## V. CONCLUSION

In this paper, a new method for automatic identification of epileptic seizures is presented using SRC and proposed dictionary learning. In the proposed method, the EEG signals are used to separate 2 to 5 classes in 9 different scenarios using the dataset recorded at the University of Bonn. We achieved 100% accuracy, sensitivity and specificity for all scenarios except C-VIII, which is very promising compared to the state-of-the-art seizure detection approaches. Furthermore, it is shown that the proposed method is robust to the measurement noise of level as much as 0 dB. It is also expected that the automated system will reduce clinician's workload in detecting subtle information hidden in the large EEG data and thus save a lot of time in identifying seizures.

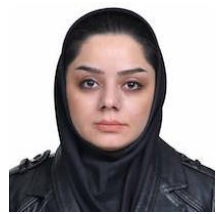

**Zohreh Mousavi** received B.Sc. degree in Mechanical Engineering from University of Yasuj, Yasuj, Iran, in 2012. and she received M.Sc. degree in Mechanical Engineering from University of Vali Asr Rafsanjan, Kerman, Iran, in 2015. She is currently Ph.D. Student in Mechanical Engineering at University of Tabriz, Tabriz, Iran. Her current research interest includes Vibration and Biomedical Signal Processing, Compressed Sensing, mechanical systems, Structural Health Monitoring (SHM), Machine Learning, Neural Networks, Deep Learning.

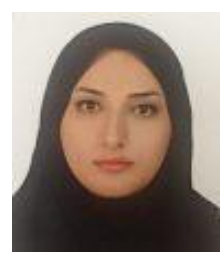

**Azra Delpak** received Medical Doctor degree from the Faculty of Medicine, Urmia University of Medical Sciences, Urmia, Iran in 2014. She is currently working as a resident in General Surgery Department, Urmia University of Medical Sciences. Her research interests consists of cognitive neuroscience, seizure, stroke, multiple sclerosis and Alzheimer's disease.

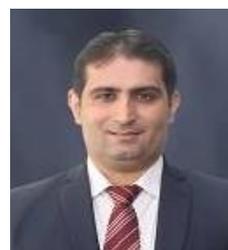

**Ali Farzamnia** received the B.Eng. degree in telecommunication engineering from the Islamic Azad University of Urmia, Urmia, Iran, in 2005, the M.Sc. degree in telecommunication engineering from the University of Tabriz, Tabriz, Iran, in 2008, and the Ph.D. degree in electrical engineering (telecommunication engineering) from Universiti Teknologi Malaysia, Johor Bahru, Malaysia, in 2014.,He is currently a Senior Lecturer with the Faculty of Engineering, Universiti Malaysia Sabah, Sabah, Malaysia. His current research interests include signal processing, network coding, and information theory.

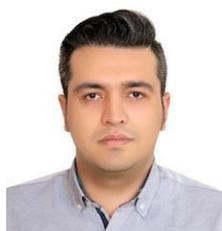

**Sobhan Sheykhivand** received B.Sc degree in Aviation-Fighter Pilot from University of Shahid Satari, Tehran, Iran in 2014 and he received B.Sc. degree in Electronic Engineering from Islamic Azad University of Urmia, Urmia, Iran, in 2016. and he received M.Sc. degree in Biomedical Engineering from University of Tabriz, Tabriz, Iran, in 2018. He is currently PhD Student in Biomedical Engineering at University of Tabriz, Tabriz, Iran. His current research interest includes biomedical signal processing, data compression and compressed sensing.

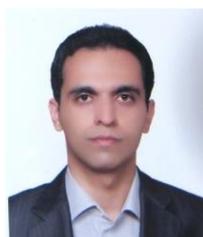

**Tohid Yousefi Rezaii** received B.Sc., M.Sc., and Ph.D. degrees all in Electrical Engineering (Communication) from University of Tabriz, Tabriz, Iran, in 2006, 2008 and 2012, respectively. He is currently with Faculty of Electrical and Computer Engineering, University of Tabriz, Tabriz, Iran. His current research interests include biomedical signal processing, data compression, compressed sensing, statistical signal processing, pattern recognitionstatistical learning and adaptive filters.Physics